\documentclass[11pt,a4paper]{scrartcl}

\usepackage[utf8]{inputenc}
\usepackage[english]{babel}           
\usepackage{amsfonts, amsmath, amsthm, amssymb, dsfont}
\usepackage{graphicx}
\usepackage{epstopdf}
\usepackage{color}
\usepackage{hyperref}

\usepackage{lineno}

\usepackage{natbib}

\newcommand{\s}{\footnotesize}  %controll the font size of figure labels
\newcommand{\xs}{\scriptsize} %controll the font size of axis tics
\newcommand{\hl}[1]{\setlength{\fboxsep}{2.0pt}\colorbox{white}{#1}} %highlights figure labels with white background

\title{Generalized modulation theory for strongly nonlinear gravity waves in a compressible atmosphere}
\author{Mark Schlutow\footnote{mark.schlutow@fu-berlin.de}\\ 
\s Institut f\"ur Mathematik, Freie Universit\"at Berlin, Germany \\
and\\
Erik Wahlén\\
\s Centre for Mathematical Sciences, Lund University, Sweden}
\date{}

\begin{document}

	\maketitle

	\begin{abstract}
		This study investigates strongly nonlinear gravity waves in the compressible atmosphere from the Earth's surface to the deep atmosphere.
These waves are effectively described by Grimshaw's dissipative modulation equations which provide the basis
for finding stationary solutions such as mountain lee waves and testing their stability in an analytic fashion.
Assuming energetically consistent boundary and far-field conditions,
that is no energy flux through the surface, free-slip boundary, and finite total energy,
general wave solutions are derived and illustrated in terms of realistic background fields.
These assumptions also imply that the wave-Reynolds number must become less than unity above a certain height.
The modulational stability of admissible, both non-hydrostatic and hydrostatic, waves is examined. 
It turns out that, when accounting for the self-induced mean flow, 
the wave-Froude number
%, as defined by the stratification times mountain height divided by horizontal background wind, 
has a resonance condition. 
If it becomes $1/\sqrt{2}$, then the wave destabilizes due to perturbations from the essential spectrum of the linearized modulation equations.
However, if the horizontal wavelength is large enough, waves overturn before they can reach the modulational stability condition.
	\end{abstract}

	\section{Introduction}
%------------------------------------------------------------------------------------	

% general introduction about gravity waves
Gravity waves are an omnipresent oscillation mode in the atmosphere.
They redistribute energy vertically but also laterally and thereby affect 
the dynamics relevant for weather and climate prediction \citep{Fritts2003,Becker2012}.
Usually excited in the troposphere, gravity waves may persist deep into the upper layers of the atmosphere \citep{Fritts2016,Fritts2018,Fritts2019}
where they interact with the mean flow.
They exert drag onto the horizontal mean-flow momentum, produce heat when dissipating \citep{Becker2004}, 
and cause increased mixing of tracer constituents such as green-house gases \citep{Schlutow2014}.

% difference between linear, weakly nonlinear and nonlinear models
The dominant excitation mechanism is background flow over mountain ranges resulting in mountain lee waves 
that can be considered stationary as their horizontal phase speed is exactly opposite to the mean-flow horizontal wind.
In the troposphere wave amplitudes are often small such that linear wave theory is applicable \citep{Eliassen1961,Putz2019}.
When mountain lee waves extend into the higher layers they get anelastically amplified due to the decreasing background density
which is an effect of the compressibility of the atmosphere.
Amplitudes get indeed so large that linear theory becomes invalid.

% modulational stability of weakly nonlinear and nonlinear Boussinesq waves
Weakly nonlinear theory for gravity waves was studied by \cite{Whitham1974a,Grimshaw1977,Sutherland2001,Sutherland2006,Tabaei2007a}.
Here, nonlinear effects such as Doppler shift of the frequency and interaction with the mean flow appear as higher-order corrections to the linear model.
In the asymptotic limit the model therefore approaches linear theory.
An important effect of the weak nonlinearity is the occurrence of modulational instabilities:
plane non-hydrostatic Boussinesq waves become modulationally unstable if the second derivative 
of the dispersion relation with respect to the wavenumber becomes negative.
\cite{Schlutow2018} showed that this holds true even for strongly nonlinear waves of the same kind.
In the strongly nonlinear theory, Doppler shift and wave-mean-flow interaction appear to leading order
such that the amplitude is finite in the asymptotic limit.
However, Boussinesq theory does not account for the anelastic amplification.
Furthermore, the plane waves extend to the infinities experiencing no lower boundary conditions 
and no dissipative effects.
In \cite{Schlutow2019}, a particular flow regime, where anelastic amplification 
and dissipative forces are exactly balanced, was investigated with respect to modulational stability.

This study aims to generalize the modulation theory incorporating strong nonlinearity, 
anelastic amplification, dissipative damping and lower boundary conditions
in a comprehensive fashion.
Pioneering work on the modulation theory of strongly nonlinear gravity waves 
was accomplished by \cite{Grimshaw1972,Grimshaw1974}.

% structure of the paper
In section \ref{sec:modeq} of this paper we will introduce Grimshaw's dissipative modulation equations 
as our governing equations and link them to the asymptotic solution to the compressible Navier-Stokes equations.
Boundary conditions and limit behavior as derived from physical arguments will be shown in sections \ref{sec:bc} and \ref{sec:limit}.
After the introduction of the antitriptic flow assumption in section \ref{sec:antitriptic},
stationary solutions will be found and illustrated in terms of observational data in section \ref{sec:statio}.
In section \ref{sec:stab}, the modulational stability of the stationary solution will be investigated
followed by some concluding remarks in section \ref{sec:conclusion}.

\section{Model equations}
\label{sec:modeq}

%Grimshaw's dissipative modulation equations
The foundation for our investigations is established by Grimshaw's dimensionless dissipative modulation equations
for horizontally periodic two-dimensional gravity waves modulated only in the vertical ($Z$-)direction
\begin{subequations}
	\label{eq:modeq}
	\begin{align}
		\label{eq:modeq_kz}		
		\frac{\partial k_z}{\partial T}+\frac{\partial}{\partial Z}(\hat\omega+k_xu)&=0\\
		\label{eq:modeq_a}	
		\rho\frac{\partial a}{\partial T}+\frac{\partial}{\partial Z}( \hat\omega'\,\rho a)&=-\Lambda |\pmb{k}|^2\rho a\\
		\label{eq:modeq_u}	
		\rho\frac{\partial u}{\partial T}+\frac{\partial}{\partial Z}(\hat\omega'\,k_x\rho a)&=-\frac{\partial p}{\partial X}
	\end{align}
\end{subequations}
in the domain $Z\in[0,\infty)$, from the Earth's surface into the deep atmosphere \citep{Grimshaw1974,Achatz2010}.
This set of coupled nonlinear partial differential equations governs the evolution of 
vertical wavenumber $k_z$, wave action density $\mathcal{A}=\rho a$, and mean-flow horizontal wind $u$.
The horizontal wavenumber $k_x$ is a constant being, without loss of generality, positive.
Intrinsic frequency $\hat{\omega}$ and linear vertical group velocity $\hat{\omega}'$ are functions of $k_z$.
They are determined by the dispersion relation for either hydrostatic ($h$) or non-hydrostatic ($nh$) gravity waves
\begin{align}
	\label{eq:dispersion}
	\hat{\omega}(k_z)=
	\begin{cases}
		Nk_x/|\pmb{k}|,~& (nh),\\
		Nk_x/|k_z|,~& (h),
	\end{cases}
	\quad\text{and}\quad
	\hat{\omega}'(k_z)=
	\begin{cases}
		-Nk_xk_z/|\pmb{k}|^3,~& (nh),\\
		-Nk_xk_z/|k_z|^3,~& (h),
	\end{cases}
\end{align}
with $|\pmb{k}|^2=k_x^2+k_z^2$.
Three coefficients appear being functions of $Z$ only: $\Lambda$, $\rho$ and $N$. 
They denote the kinematic viscosity, background density and the Brunt-V\"ais\"ala frequency, respectively.
$\rho$ and $N$ are computed assuming a hydrostatic atmosphere being an ideal gas.
Given a background temperature profile $\mathcal{T}(Z)$ we obtain
\begin{subequations}
	\begin{align}
		\label{eq:bvf}
		N^2(Z) &=\frac{1}{\mathcal{T}}\left(\frac{d\mathcal{T}}{dZ}+1\right),\\
		\label{eq:rho}
		\rho(Z) &=\exp\left(\int_0^Z\eta(Z')\,\mathrm{d}Z'\right)\quad\text{with}\quad
		\eta(Z)=-\frac{1}{\mathcal{T}}\left(\frac{d\mathcal{T}}{dZ}+\frac{1}{\kappa}\right)
	\end{align}
\end{subequations}
where $\kappa=(\gamma-1)/\gamma=2/7$ and $\gamma$ the heat capacity ratio for ideal gases. 
The height profile of the kinematic viscosity $\Lambda$ is determined by the sum of turbulent and molecular viscosity. 
$\Lambda$ is negligible at the surface and in the lower atmosphere. 
At a certain height, however, amplitudes of gravity waves become typically so large due to the anelasitic amplification 
that turbulence is produced by small-scale instabilities which increases $\Lambda$. 
Even higher in the atmosphere, about 100\,km, molecular viscosity starts to dominate becoming orders of magnitude larger than the turbulent viscosity.
Eventually in the highest layer, the thermosphere, molecular dissipation damps effectively every wave motion.

% Summary of physiscs
In Whitham's modulation theory \citep{Whitham1974a}, equation \eqref{eq:modeq_kz} represents conservation of waves.
The second equation \eqref{eq:modeq_a} gives conservation of wave action plus a sink due to dissipation.
Finally, \eqref{eq:modeq_u} describes the acceleration of mean-flow horizontal momentum
due to horizontal pseudo-momentum flux convergence and the horizontal gradient of mean-flow kinematic pressure $p$.
The latter is unspecified at the moment and needs further assumptions to close the system.

% Asymptotic solution to the compressible Navier-Stokes equation
All prognostic fields generally depend on the compressed dimensionless coordinates
\begin{align}
	(X,\,Z,\,T)&=(\varepsilon^\alpha x,\,\varepsilon z,\,\varepsilon^\alpha t)
	\quad\text{where}\quad
	\alpha=
	\begin{cases}
		1,~& (nh),\\
		2,~& (h),
	\end{cases}
\end{align}
that originate from a rescaling of the dimensionless coordinates
\begin{align}
	(x,\,z,\,t)&=(x^\ast/L_r,\,z^\ast/L_r,\,t^\ast/t_r).
\end{align}
The dimensionless coordinates are obtained by the non-dimensionalization with the reference length scale $L_r$ and reference time scale $t_r$. 
Note that throughout this work, dimensioned variables are labeled with an asterisk.  
Scale separation between the slow variation of the modulational fields and the rapidly oscillating wave field is determined by $0<\varepsilon\ll 1$.
The modulation equations \eqref{eq:modeq} can be derived from the compressible Navier-Stokes equations (NSE)
in an asymptotic fashion as $\varepsilon\rightarrow 0$ which is shown in a comprehensive manner in \citet{Schlutow2017b}.
For this derivation, a distinguished limit on the dimensionless NSE is assumed that 
maps $\varepsilon$ to the Mach, Froude, Reynolds and Prandtl number, individually.
The background field is separated from the perturbing wave field by a scaling appropriate 
for either hydrostatic or non-hydrostatic waves obtained from linear theory.
Next, the state vector $\pmb{U}=(\pmb{v},\,\beta,\,\phi)^\mathrm{T}$ of the NSE is expanded 
in terms of a nonlinear Wentzel-Kramers-Brillouin (WKB) ansatz 
\begin{align}
	\label{eq:asym_expa}
	\pmb{U}(x,z,t;\varepsilon)=\pmb{U}_{0,0}(X,Z,T)+\left(\pmb{U}_{0,1}(X,Z,T)e^{i\Phi(X,Z,T)/\varepsilon}
	+\mathrm{c.c.}\right)+\mathrm{h.h.}+\mathit{O}(\varepsilon)
\end{align}
where c.c. stands for the complex conjugate and h.h. for higher harmonics.
Here, $\pmb{v}$, $N\beta=b$ and $\phi$ denote the two-dimensional velocity vector, buoyancy force and kinematic pressure, respectively.
By construction, the WKB ansatz is indeed a strongly nonlinear approach---in the asymptotic limit the amplitudes are finite. In fact, the ansatz converges
to the nonlinear plane wave of Boussinesq theory which is known to be an analytical
solution.
This ansatz is substituted into the NSE and terms are ordered in powers of $\varepsilon$ 
and harmonics, i.e. integer multiples of the phase $\Phi$.
Then, the leading-order solution of the NSE as $\varepsilon\rightarrow 0$ is given in terms of 
the modulation fields governed by \eqref{eq:modeq} as
\begin{subequations}
	\begin{align}
		\left(\frac{\partial\Phi}{\partial X},\,\frac{\partial\Phi}{\partial Z}\right)^\mathrm{T}&=\pmb{k},\\
		\label{eq:def_omega}
		-\frac{\partial\Phi}{\partial T}&=\hat\omega+k_xu,\\
		\pmb{U}_{0,0}&=(u,\,0,\,0,\,p)^\mathrm{T},\\
		\label{eq:u01}
		\pmb{U}_{0,1}&=\mathcal{B}\,\pmb{U}^\dagger
	\end{align}
\end{subequations}
where 
\begin{align}
	\label{eq:pol_vec}
	\pmb{U}^\dagger=\left(-i\frac{k_z}{k_x}\frac{\hat{\omega}}{N},\,i\frac{\hat{\omega}}{N},\,
			1,\,-i\frac{k_z}{k_x^2}\frac{\hat{\omega}^2}{N}\right)^\mathrm{T}
\end{align}
represents the polarization vector and $\mathcal{B}=\sqrt{\hat\omega a/2}$ the amplitude.

Furthermore, the modulation equations exhibit a total energy density
being the sum of mean-flow kinetic energy density and wave energy density,
\begin{align}
	\label{eq:erg}
	\rho e=\frac{1}{2}\rho u^2+\rho a\hat{\omega}.
\end{align}
It evolves in time as governed by
\begin{align}
	\label{eq:loc_erg}
	\rho\frac{\partial e}{\partial T}+\frac{\partial}{\partial Z}\left(\hat{\omega}'\rho a(\hat{\omega}+k_xu)\right)+\frac{\partial}{\partial X}(pu)
	=-\hat{\omega}\Lambda|\pmb{k}|^2\rho a.
\end{align}
In conclusion, \eqref{eq:erg} is a locally conserved quantity in the inviscid limit, i.e. $\Lambda\rightarrow 0$.
%and when the horizontal gradient of mean-flow kinematic pressure vanishes.

\section{Surface boundary condition}
\label{sec:bc}
%------------------------------------------------------------------------------------
%
The leading-order asymptotic solution to the NSE as defined by \eqref{eq:asym_expa} governed by the modulation equations
already determines the shape of the anticipated wave solution. 
However, it has some degrees of freedom to specify physically motivated surface boundary conditions.
Given a mean-flow horizontal wind, three additional conditions are needed 
to set the horizontal and vertical wavenumber as well as
the specific wave action density at $Z=0$.
We will assume that close to the surface the viscosity is negligible. 
Thus, a free-slip boundary is justified. 
Let us carve a mountain to the shape of the lowest stream line.
Then, this particular, prototypical mountain will define two of the surface boundary conditions.
For the third, we will assume that there shall be no energy flux through the boundary and therefore  
the extrinsic frequency will be zero at the mountain.

\subsection{No-energy-flux boundary condition}

According to \eqref{eq:loc_erg}, if no energy flux through  the boundary 
but finite wave action and wavenumber at the boundary are presumed, then
\begin{align}
	\label{eq:bc2}
	\hat{\omega}+k_xu=0\quad\text{at}\quad Z=0
\end{align}
must hold which coincides with the absence of wavenumber flux (cf. \eqref{eq:modeq_kz}).
Let us remark that therefore $u<0$ as $k_x>0$ and $\hat{\omega}>0$ at $Z=0$.

\subsection{Free-slip boundary condition: carving a mountain to the wave}

\begin{figure*}
	\begin{center}
		\input{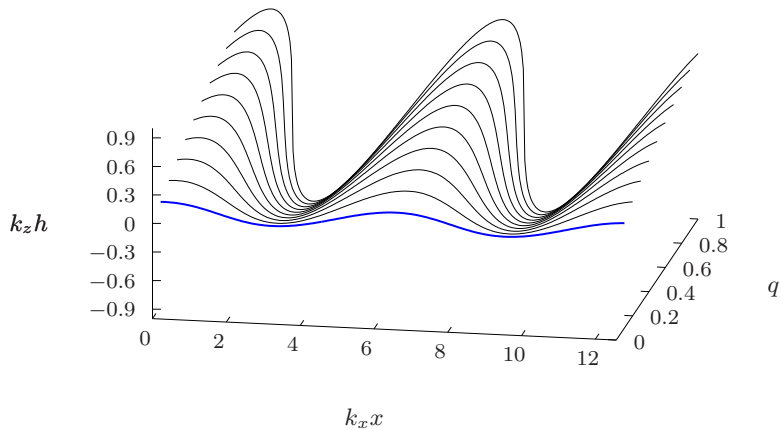}	
	\end{center}
	\caption{Stream lines as fixed points of $f$ for different $q$.}
	\label{fig:streamlines}
\end{figure*}

To leading order the solution as given by \eqref{eq:asym_expa} is solenoidal and therefore entirely
determined by a stream function.
From the polarization vector \eqref{eq:pol_vec}, the stream function can be written in terms of the fast variables as
\begin{align}
	\Psi(x,z)=uz-2\mathcal{B}\frac{\hat{\omega}}{Nk_x}\cos(k_xx+k_zz)+\mathit{O}(\varepsilon).
\end{align}
%	such that $\mathbf{v}=\nabla_\perp\Psi+\mathit{O}(\varepsilon)$.
Stream functions of solenoidal flows are constant on stream lines $\bigl(x,h(x)\bigr)$
which can be formulated mathematically by
\begin{align}
	\label{eq:stream_line}
	\Psi(x,h(x))=const.
\end{align}
Considering the stream line in an $O(\varepsilon)$-neighborhood of the boundary $Z=0$,
\eqref{eq:stream_line} defines a parametrization of $h$ implicitly via
\begin{align}
	\label{eq:def_oro}
	h(x)=\frac{2\mathcal{B}}{u}\frac{\hat{\omega}}{Nk_x}\cos(k_xx+k_zh(x))+\mathit{O}(\varepsilon).
\end{align}
Here, we have essentially presented the boundary condition of \citet[][Eq. 7]{Lilly1979},
who define the condition in terms of the vertical displacement of an air parcel $\delta=-\Psi/u$,
but in the framework of Grimshaw's dissipative modulation equations.
Note that in the asymptotic limit $\varepsilon\rightarrow 0$, \eqref{eq:def_oro} is of the form 
\begin{align}
	y=f(y)
\end{align}
with $y(\xi)=k_zh$, $\xi=k_xx$ and $f=q\cos(\xi+y)$ 
where 
\begin{align}
	\label{eq:nondimensional_mountain_height}
	q=\frac{2\mathcal{B}}{u}\frac{\hat{\omega}k_z}{Nk_x}.
\end{align}
Thus, the stream line can be interpreted as a fixed point of $f$.
For every fixed $\xi$ the differentiable function $f$ of $y$ has a Lipschitz constant
\begin{align}
	L=\left|\sup_y f'(y)\right|=|q|.
\end{align}
Banach's fixed point theorem states if $L=|q|<1$, then $f$ is a contraction and hence has a fixed point.
The results	of a fixed-point iteration $y_{n+1}=F(y_n)$ with $y_0=0$ are plotted in figure \ref{fig:streamlines}.
It can be observed how the stream lines steepen when $q$ is increased
whereas for small values the stream lines approximate the sinusoidal (linear) profile.
Exploiting the polarization \eqref{eq:pol_vec}, we can write
\begin{align}
	q=\frac{u'}{|u|},\quad u'=2|u_{0,1}|=2\mathcal{B}\frac{\hat{\omega}|k_z|}{Nk_x}
\end{align}
which provides a measure for nonlinearity.
If it is small one may assume linear wave excitation. 
The factor 2 appears simply because of the definition of amplitude due to \eqref{eq:asym_expa}.
Also, using polarization \eqref{eq:pol_vec} and no-energy-flux \eqref{eq:bc2}
in combination with the convergence condition for the fixed-point iteration, 
we obtain
\begin{align}
	\label{eq:stat_stab}
	N^2>|k_z|b',\quad b'=2\mathcal{B}N
	\end{align}
which confirms the classical condition for static stability, 
lines of constant potential temperature must not overturn \citep{Boloni2016}.

A boundary condition close to the surface but sufficiently 
far away to be considered free-slip assuming $\Lambda(0)=0$
can be determined by a periodic mountain ridge 
with period $\mathcal{P}$ and maximum mountain height $H_m$
which we consider to be given constants hereinafter.
In terms of the nonlinearity parameter \eqref{eq:nondimensional_mountain_height} 
and the no-energy-flux condition \eqref{eq:bc2}, 
we obtain that $k_x=2\pi/\mathcal{P}$ and
\begin{align}
	\label{eq:bc1}
	H_m=\frac{2\mathcal{B}}{N}\quad\text{at}\quad Z=0.
\end{align}

The combined boundary condition from \eqref{eq:bc1} and \eqref{eq:bc2} may be written in vector form as
\begin{align}
	\label{eq:bc}
	\pmb{B}(\pmb{y})=
	\begin{pmatrix}
		\hat{\omega}+k_xu\\
		H_m^2N^2-2a\hat{\omega}
	\end{pmatrix}
	=0\quad\text{at}\quad Z=0.
\end{align}

\section{Far-field condition - limit behavior}
\label{sec:limit}

Additionally to the lower boundary, the limit of the solution as $Z\rightarrow\infty$
must be specified in order to obtain a physical wave solution. 
For this argument, we exploit the global energy as derived from \eqref{eq:erg} being
\begin{align}
	\label{eq:glob_erg}
	\int_0^\infty\rho e\,\mathrm{d}Z=\int_0^\infty\frac{1}{2}\rho u^2\,\mathrm{d}Z
	+\int_0^\infty\rho a\hat{\omega}\,\mathrm{d}Z.
\end{align}
A physical wave must be such that the global energy is finite.

In the thermosphere, temperature approaches an equilibrium $T_\infty$ as altitude increases \citep{Rawer1984}. 
Hence, it is safe to assume that $N\rightarrow N_\infty$ and $\eta\rightarrow \eta_\infty<0$ 
as $Z\rightarrow\infty$ (cf. \eqref{eq:bvf} and \eqref{eq:rho}).
Due to the high viscosity in the thermosphere and above, we can also assume that  $u,a\rightarrow 0$ as $Z\rightarrow\infty$.
These assumptions suffice for the integrals in \eqref{eq:glob_erg} to converge 
and also that the energy flux in \eqref{eq:loc_erg} vanishes.
Additionally, we assume that the background variables are exponentially asymptotic, 
i.e. they converge even when multiplied by an exponentially increasing function \citep[][p 40]{Kapitula2013}.

\section{The antitriptic flow assumption and momentum conservation}
\label{sec:antitriptic}

In the inviscid limit the modulation equations \eqref{eq:modeq} assume stationary solutions
where ${\partial p/\partial X=0}$ which can be computed analytically by a formula of \citet[][Eq. 5.20]{Schlutow2017b}.
When we multiply \eqref{eq:modeq_a} by $k_x$ and subtract \eqref{eq:modeq_u},
we obtain the evolution equation for total momentum density,
\begin{align}
	\label{eq:stat_cond}
	\rho\frac{\partial }{\partial T}(k_xa-u)=\frac{\partial p}{\partial X}-k_x\Lambda |\pmb{k}|^2\rho a.
\end{align}
Thus, to be consistent with the inviscid limit, the dissipative modulation equations assume stationary solutions %only 
if the right hand side of \eqref{eq:stat_cond} vanishes which provides eventually a closure 
for the horizontal gradient of mean-flow kinematic pressure,
\begin{align}
	\label{eq:mfpg}
	\frac{\partial p}{\partial X}=\Lambda |\pmb{k}|^2\rho k_xa.
\end{align}
This result implicates that the mean-flow horizontal kinematic pressure gradient 
balances the dissipation of horizontal pseudo-momentum, $\rho k_xa$.
Thereby, total momentum is locally conserved. 
Also, let us remark that there is no vertical momentum flux.
Gravity waves carry only pseudo-momentum \cite[cf.][]{Mcintyre1981}.

A flow configuration where pressure gradient balances viscous forces is referred to as antitriptic flow in the literature \citep{Jeffreys1922}.
Under this antitriptic flow assumption \eqref{eq:mfpg} the modulation equations degenerate
as we can integrate \eqref{eq:stat_cond} with respect to time to obtain the mean-flow horizontal wind, 
\begin{align}
	\label{eq:degeneration}
	u(Z,T) = k_xa(Z,T)+\bar{U}(Z),
\end{align}
reducing to a diagnostic variable.
We call $\bar{U}$ the background horizontal wind since it is time-independent 
and we emphasize that in the absence of a wave ($a=0$), 
mean-flow and background horizontal wind coincide.
The difference between mean-flow and background horizontal wind can be identified as specific horizontal pseudo-momentum.
Note that in weakly nonlinear wave theory, 
the pseudo-momentum is of higher order and would not appear in a leading-order equation as shown here.

% Vector form
Substituting \eqref{eq:degeneration} into the governing equations \eqref{eq:modeq}
reduces the dimension of the system. We may reformulate the reduced system in vector form
\begin{align}
	\label{eq:modeq_vec}
	\frac{\partial \pmb{y}}{\partial T}+\frac{\partial \pmb{F}(\pmb{y})}{\partial Z}=\pmb{G}(\pmb{y})
\end{align}
where we call $\pmb{y}=(k_z,\,a)^\mathrm{T}:[0,\infty)^2\rightarrow\mathbb{R}^2$ the prognostic state vector.
The nonlinear flux and inhomogeneity are determined by
\begin{subequations}
	\begin{align}
		\pmb{F}&=\left(\hat{\omega}+k_xu,\,\hat{\omega}'\rho a\right)^\mathrm{T},\\
		\pmb{G}&=\left(0, -(\Lambda|\pmb{k}|+\eta\hat{\omega}')a\right)^\mathrm{T}.
	\end{align} 
\end{subequations}

\section{Stationary solution}
\label{sec:statio}

This section will examine stationary solutions to Grimshaw's dissipative modulation equations.
They describe typical mountain lee waves which are excited by a background flow over a mountain. 

\subsection{Derivation for stationary waves}

A stationary solution $\pmb{y}=\pmb{Y}(Z)=(K_z,\,A)^\mathrm{T}$ must fulfill
\begin{align}
	\label{eq:statio}
	\frac{\partial \pmb{F}(\pmb{Y})}{\partial Z}=\pmb{G}(\pmb{Y}).
\end{align}
Note that we label the stationary solution with capital letters.
The first component of \eqref{eq:statio} can readily be integrated, so
\begin{align}
	\label{eq:statio_om}
	const.&=\Omega=\hat{\Omega}+K_xU.
\end{align}
with $U(A)=K_xA+\bar{U}$ and $\hat{\Omega}^{(n)}=\hat{\omega}^{(n)}(K_z)$.
We can solve \eqref{eq:statio_om} explicitly exploiting \eqref{eq:dispersion} for
\begin{align}
	\label{eq:statio_kz}
	K_z(A)=
	\begin{cases}
		-K_x\sqrt{N^2/(K_xU(A)-\Omega)^2-1},\quad&(nh),\\
		K_xN/(K_xU(A)-\Omega),\quad&(h).
	\end{cases}
\end{align}
We want to point out that in the non-hydrostatic case, 
the vertical wavenumber may become imaginary which is referred to as evanescence
since the ansatz \eqref{eq:asym_expa} would switch from an oscillatory to an exponential behavior.
In fact vertically evanescent waves are observed in the atmosphere and can be modeled by linear theory.
However, the nonlinear WKB theory becomes invalid for imaginary wavenumbers.
The second component of \eqref{eq:statio} becomes an explicit, non-autonomous, scalar, ordinary differential equation
when we insert \eqref{eq:statio_kz},
\begin{subequations}
	\label{eq:statio_all}
	\begin{align}
		\label{eq:statio_a}
		\frac{\partial A}{\partial Z}&=\Gamma(A,Z)\,A\quad\text{in}\quad(0,\infty),\\
		\label{eq:gamma}
		\Gamma&=\frac{\eta(\mathit{Re}_\mathrm{wave}^{-1}-1)
		-(1-\alpha\hat{\Omega})\partial_Z\ln(N)
		+\alpha K_x\partial_Z\bar{U}}{1-K_x^2\alpha A},\\
		\alpha&=\frac{\hat{\Omega}''}{\hat{\Omega}^{\prime 2}},\\
		\label{eq:wavere}
		\mathit{Re}_\mathrm{wave}&=\frac{|\eta|\hat{\Omega}'}{\Lambda|\pmb{K}|^2}.
	\end{align}
\end{subequations}
Definition \eqref{eq:wavere} can be interpreted as a wave-Reynolds number. 
It was introduced and discussed by \citet{Schlutow2019} and measures, roughly speaking, the damping of wave amplitude.

From the boundary condition \eqref{eq:bc}, $\pmb{B}(\pmb{Y})=0$ at $Z=0$, we obtain
\begin{align}
	\Omega&=0,\\
	\label{eq:bca}
	A^\pm&=\frac{|\bar{U}|}{2K_x}\left(1\mp\sqrt{1-2\mathit{Fr}_\mathrm{wave}^2}\right)\quad\text{at}\quad Z=0
\end{align}
where
\begin{align}
	\label{eq:wavefroude}
	\mathit{Fr}_\mathrm{wave}=\left.\frac{H_mN}{|\bar{U}|}\right|_{Z=0}\leq\frac{1}{\sqrt{2}}.
\end{align}
We can rule out $A^-$ immediately as non-physical solution as there must be no wave if $H_m=0$. 
The constant $\mathit{Fr}_\mathrm{wave}$ possesses an upper bound in order to obtain real-valued wave action.
It is often referred to as ``non-dimensional mountain height'' in the literature.
Other names are also common.
However, we follow a recent discussion by \citet{Mayer2017} who argue to call it
wave-Froude number.

For a brief discussion on the existence of solutions to \eqref{eq:statio_all} we restrict ourselves to the hydrostatic and isothermal case,
so let $N,\eta$ and $\bar{U}$ be constant.
Then, \eqref{eq:statio_a} reduces to
\begin{align}
	\label{eq:auton_statio_a}
	\frac{\partial A}{\partial Z}&=
	\eta\frac{\mathit{Re}_\mathrm{wave}^{-1}-1}{1-K_zN^{-1}K_xA}
	\,A\quad\text{in}\quad(0,\infty).
\end{align}
We can immediately observe that the denominator is always greater than one as $K_z<0$
and due to this fact, a singularity cannot be reached.
By definition $\mathit{Re}_\mathrm{wave}^{-1}=0$ at $Z=0$ and $\eta<0$
which implies that the solution will grow exponentially from the boundary.
This property is known from linear theory and is called anelastic amplification.
The viscosity $\Lambda$ increases with height and in conclusion, 
the wave-Reynolds number drops until it becomes unity.
At this point, $A$ assumes it maximum.
Physically speaking, it saturates due to turbulent damping. 
In order to meet the requirements in the deep-atmosphere limit from section \ref{sec:limit}, $A\rightarrow 0$, 
we must find
\begin{align}
	\label{eq:turbre}
	\mathit{Re}_\mathrm{wave}<1\quad\text{for }Z>Z_\mathrm{turbo}.
\end{align}
$Z_\mathrm{turbo}$ marks the turbopause, i.e. the transition zone from turbulent-dominated damping
to energy diffusion by molecular viscosity \citep{Nicolet1954}.
Due to the strong molecular dissipation above the turbopause increasing $\Lambda$, 
there can be no doubt that the bound \eqref{eq:turbre} will be reached.
But its position $Z_\mathrm{turbo}$ is hard to predict as it depends on small-scale processes which are not governed by the modulation equations anymore.
However, we will see in section \ref{sec:stab} that the actual value of $Z_\mathrm{turbo}$ has no influence on the stability of the waves.

The general case, non-isothermal and non-hydrostatic, will be investigated numerically in the following section.

\subsection{Illustrative example}

\begin{figure*}
	\begin{center}
		\input{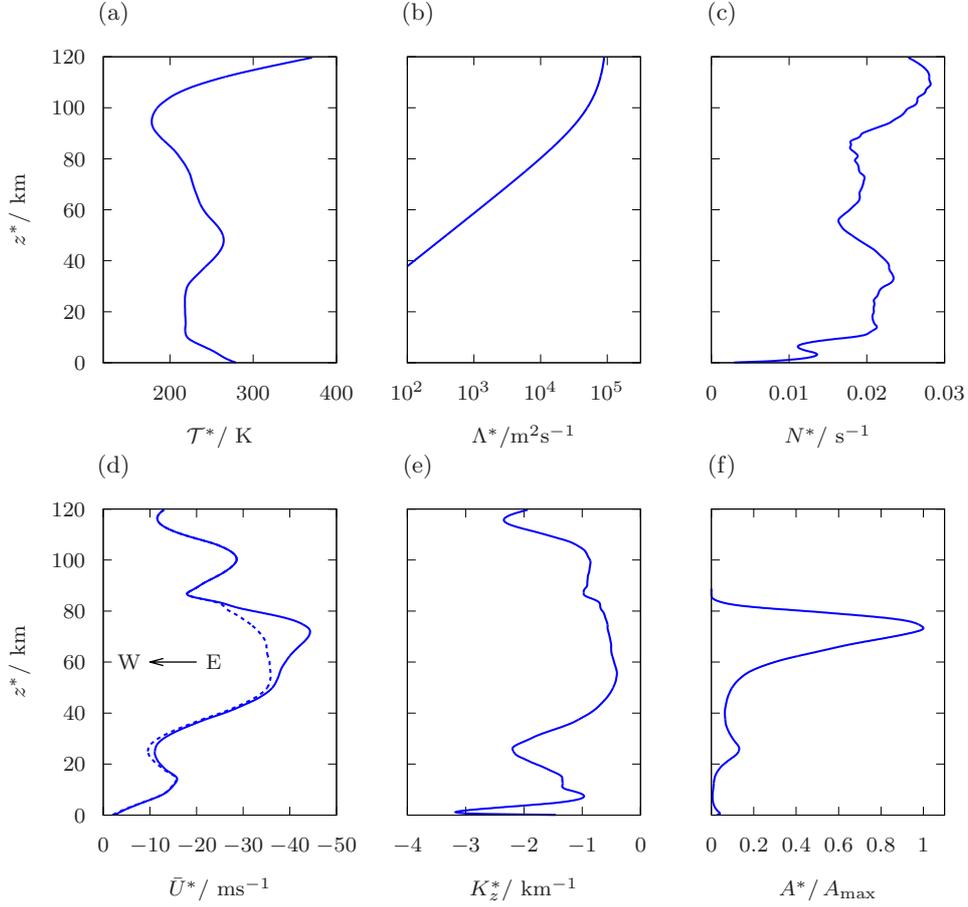}	
	\end{center}
	\caption{Illustrative example. CIRA-86 zonally averaged temperature (panel a) 
	and background horizontal wind (panel d, thick blue line) for March at 50$^\circ$\,N.
	Kinematic viscosity (panel b), Brunt-V\"ais\"al\"a frequency (panel c) computed by \eqref{eq:bvf}.
	Vertical wave number (panel e) and specific wave action density (panel f) computed by \eqref{eq:statio_kz} and \eqref{eq:statio_a}, respectively, assuming
	horizontal wavenumber $K_x^\ast = 0.2$\,km$^{-1}$.
	Wave-induced mean flow (panel d, dashed blue line).}
	\label{fig:profiles}
\end{figure*}

An illustrative example for a typical mountain wave is plotted in figure \ref{fig:profiles}. 
We took observation data from the zonal mean COSPAR International Reference Atmosphere (CIRA-86, \citet{Fleming1990})
for the dimensional temperature profile $\mathcal{T}^\ast$ (Panel a) and the dimensional background horizontal wind $\bar{U}^\ast$ (Panel d).
The values are taken for March in the northern hemisphere at 50$^\circ$ 
as here conditions for waves that extend deep into the atmosphere are optimal.
We defined the dimensional kinematic viscosity profile $\Lambda^\ast(z^\ast)\propto \tanh$
in extrapolation of \citet[][Fig. 1]{Walterscheid2011}
since viscosity is unavailable in the CIRA-86 data set.
In contrast to the common definition, we span the zonal $x$-axis from East to West due to our sign convention ($k_x>0$).
Background and wave fields are computed accordingly (see figure caption for equation references).
The data set contains important features of the atmosphere such as clearly pronounced 
troposphere, stratosphere, mesosphere and thermosphere due to the temperature inflection points.
Also, the polar and mesospheric jets are distinctly visible.
A typical wave envelope profile is generated in terms of standard numerical solvers for \eqref{eq:statio_all} (panel d).
It grows exponentially with height as the background density decreases
until viscosity starts to dominate damping the wave to disappearance.
As a final remark, we must recognize that the data set does not reach far enough into the higher atmosphere to observe the far-field behavior 
as discussed in Section \ref{sec:limit} and shown in \citet{Rawer1984}.

%------------------------------------------------------------------------------------
\section{Modulational stability of the stationary solution}
\label{sec:stab}

This section is dedicated to the stability of the stationary solution of the modulation equations.
In order to assess stability we linearize the governing equations \eqref{eq:modeq_vec} 
and the boundary condition \eqref{eq:bc}
around the stationary solution \eqref{eq:statio} and apply the ansatz for the pertubation
\begin{align}
	\pmb{y}(Z,T)=\hat{\pmb{y}}(Z)e^{\lambda T}.
\end{align}
This transforms the problem of stability into a boundary eigenvalue problem (BEVP),
\begin{subequations}
\label{eq:evp}
\begin{alignat}{4}
	\lambda\hat{\pmb y}+\frac{\partial}{\partial Z}\left[\mathrm{D}\pmb{F}(\pmb{Y})\hat{\pmb y}\right]
	&=\mathrm{D}\pmb{G}(\pmb{Y})\hat{\pmb y}\quad&\text{for}\quad&Z\in(0,\infty),\\
	\mathrm{D}\pmb{B}(\pmb{Y})\hat{\pmb{y}}&=0\quad&\text{at}\quad&Z=0
\end{alignat}
\end{subequations}
with the Jacobian matrices
\begin{subequations}
	\begin{align}
		\mathrm{D}\pmb{F}(\pmb{Y})&=
		\begin{pmatrix}
			\hat\Omega'& K_x^2\\
			\hat\Omega''A& \hat{\Omega}'
		\end{pmatrix},\\
		\mathrm{D}\pmb{G}(\pmb{Y})&=
		\begin{pmatrix}
			0& 0\\
			-\eta \hat{\Omega}''A-2\Lambda K_zA& -\eta\hat{\Omega}'-\Lambda|\pmb{K}|^2
		\end{pmatrix},\\
		\mathrm{D}\pmb{B}(\pmb{Y})&=
		\begin{pmatrix}
			\hat{\Omega}'& K_x^2\\
			-2A\hat{\Omega}'& -2\hat{\Omega}
%			-6A\hat{\Omega}^2\hat{\Omega}'& 2H_m^2N^2K_x^3U-2\hat{\Omega}^3
		\end{pmatrix}.
	\end{align}
\end{subequations}
Solving the BEVP really means to find the spectrum 
of the linear differential operator $\mathcal{L}$ defined due to \eqref{eq:evp}.
The wave is stable if there is no spectrum on the right-hand side of the complex plane.
We can decompose the spectrum into the essential (continuous) 
and the point (matrix-like) spectrum.
A comprehensive introduction in this method can be found in \citet{Sandstede2002}.
In the following sections, we will study each part of the spectrum individually.

\subsection{Essential spectrum}
\label{sec:es}

The linear operator $\mathcal{L}$ of the BEVP
can be approximated by an asymptotic differential operator having constant coefficients, $\mathcal{L}_\infty$.
Utilizing Fredholm operator theory one can prove that it 
has the same essential spectrum as the original operator \citep{Kapitula2013}.
The BEVP of the asymptotic operator can be reformulated as an initial value problem
\begin{subequations}
\begin{alignat}{4}
	\frac{\partial\hat{\pmb{y}}}{\partial Z}&=\mathbf{C}_\infty(\lambda)\hat{\pmb{y}}
	\quad&\text{for}\quad&Z\in(0,\infty),\\
	\mathrm{D}\pmb{B}(\pmb{Y})\hat{\pmb{y}}&=0\quad&\text{at}\quad&Z=0	
\end{alignat}
\end{subequations}
where the constant coefficient matrix is given by
\begin{align}
	\mathbf{C}_\infty(\lambda)&=\lim_{Z\rightarrow+\infty}\mathbf{C}(Z,\lambda),\\
	\mathbf{C}(Z,\lambda)&=\mathrm{D}\pmb{F}(\pmb{Y})^{-1}\left(\mathrm{D}\pmb{G}(\pmb{Y})
	-\frac{\partial\mathrm{D}\pmb{F}(\pmb{Y})}{\partial Z}-\lambda\right).
\end{align}
Existence of the limit is granted due to the assumptions of section \ref{sec:limit}.
The asymptotic operator $\mathcal{L}_\infty-\lambda$, and hence the original operator $\mathcal{L}-\lambda$, are Fredholm if $C_\infty$ is hyperbolic,
i.e. all its eigenvalues have non-zero real part.
We find two distinct spatial eigenvalues
\begin{subequations}
\begin{align}
	\nu_1(\lambda)&=-\frac{\lambda}{\hat{\Omega}'_\infty},\\
	\nu_2(\lambda)&=-\frac{\lambda+\eta_\infty\hat{\Omega}'_\infty+\Lambda_\infty|\pmb{K}_\infty|^2}{\hat{\Omega}'_\infty}.
\end{align}
\end{subequations}
Thus, the Morse index, which is defined as the dimension of the unstable subspace of a hyperbolic matrix, is
\begin{align}
	\label{eq:morse}
	i_\infty(\lambda)=\left\lbrace\begin{array}{lrrcl}
		0 & \text{if } & 0<&\Re(\lambda),&{}\\
		1 & \text{if } & -\eta_\infty\hat{\Omega}'_\infty-\Lambda_\infty|\mathbf{K}_\infty|^2<&\Re(\lambda)&<0,\\
		2 & \text{if } & {} &\Re(\lambda)&<-\eta_\infty\hat{\Omega}'_\infty-\Lambda_\infty|\pmb{K}_\infty|^2.
	\end{array}\right.
\end{align}
Note that $-\eta_\infty\hat{\Omega}'_\infty-\Lambda_\infty|\pmb{K}_\infty|^2<0$ due to \eqref{eq:turbre}.
On lines in the complex plane where $\Re(\lambda)=0$ and $\Re(\lambda)=-\eta_\infty\hat{\Omega}'_\infty-\Lambda_\infty|\pmb{K}_\infty|^2$
the matrix $C_\infty$ is not hyperbolic and hence the operator is not Fredholm.
The Fredholm index tells us where the essential spectrum lies.
According to \cite[][their formula 1.10]{Ben-Artzi1993} and also \cite[][p 391]{Gohberg1990}, 
it can be written as
\begin{align}
	\operatorname{ind}=\dim\bigl(\ker\mathrm{D}\pmb{B}(\pmb{Y}_0)\bigr)-i_\infty(\lambda)
\end{align}
where $\pmb{Y}(0)=\pmb{Y}_0$.
The essential spectrum is the set of $\lambda$'s 
for which the operator $\mathcal{L}-\lambda$ is Fredholm but ind $\neq 0$ or it is not Fredholm. 
The point spectrum, on the other hand, lies where the operator is Fredholm and ind $=0$ but the operator is not invertible.
We will investigate the point spectrum in section \ref{sec:pts}.
Having a closer look on \eqref{eq:morse}, it turns out 
that 
\begin{align}
	\dim\bigl(\ker\mathrm{D}\pmb{B}(\pmb{Y}_0)\bigr)=0
\end{align}		
must be true in order to obtain a stable essential spectrum,
i.e. no essential spectrum on the right hand side of the complex plane.
In particular, if the kernel is non-empty, the waves are unstable and the operator is even ill-posed
as the complete right plane is in the essential spectrum.
The criterion can be rephrased: stable waves necessitate Dirichlet boundary conditions.
This is violated and hence the wave destabilizes due to perturbations from the essential spectrum if
\begin{align}
	\frac{H_mN}{|U|}=\sqrt{2}\quad\text{at }Z=0
\end{align}
which can be called the gross wave-Froude number. 
This modulational instability criterion is not determined 
by an inequality like most fluid dynamical stability criteria. 
In fact the instability condition has to be fulfilled by equality. 
It is therefore more similar to a catastrophic resonance condition.

\subsection{Point spectrum}
\label{sec:pts}

In this section we will prove the non-existence of unstable point spectrum.
Let us assume a stable essential spectrum 
and the existence of an unstable eigenvalue.
For this eigenvalue the Fredholm index is zero 
and hence it may belong to the point spectrum.
Then, the eigenfunction solves the associated ODE 
\begin{align}
	\frac{\partial\hat{\pmb{y}}}{\partial Z}=\mathbf{C}(Z,\lambda)\hat{\pmb{y}}.
\end{align}
By assumption, the essential spectrum is stable and hence the kernel of the Jacobian of the boundary condition is empty 
or, in other words, we get a Dirichlet boundary condition, $\hat{\pmb{y}}=0$ at $Z=0$.
The Dirichlet boundary is the initial condition for the ODE
which then assumes the trivial solution, so there is no eigenfunction.
This contradiction completes the proof that there is no unstable point spectrum.
In conclusion, modulational instabilities, given that the resonance condition is true, 
originate from the essential spectrum.

\subsection{Summary}

\begin{figure*}
	\begin{center}
		\input{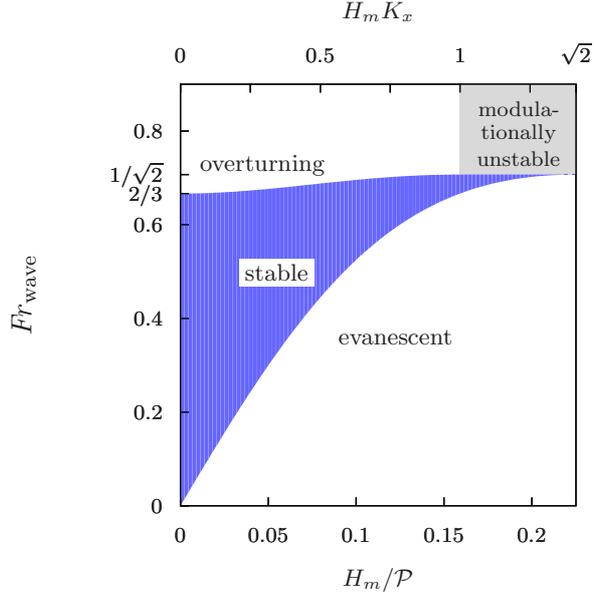}	
	\end{center}
	\caption{Admissible wave-Froude number as function of height-period ratio for non-hydrostatic waves.
	For hydrostatic waves, it is the same result but $\mathcal{P}\rightarrow\infty$ ($K_x=0$).
	Only values in the blue area correspond to wave solutions.}
	\label{fig:wave_froude}
\end{figure*}

We will summarize all results regarding stability of the stationary solution in this section.
The gross wave-Froude number depending on the wave itself due to the induced mean flow 
is specified in terms of \eqref{eq:bca} and \eqref{eq:wavefroude} by
\begin{align}
	\label{eq:growavefr}
	\frac{H_mN}{|U|}=\frac{2\mathit{Fr}_\mathrm{wave}}{\sqrt{1-2\mathit{Fr}_\mathrm{wave}^2}+1}\quad\text{at }Z=0.
\end{align}
In conclusion, stationary waves extending from the surface to the deep atmosphere, 
that experience 
\begin{align}
	\mathit{Re}_\mathrm{wave}<1\quad\text{above }Z_\mathrm{turbo},
\end{align}  
are 
\begin{itemize}
	\item non-evanescent and statically stable if
	\begin{align}
		\begin{cases}
		0,~& (h),\\
%		2\pi H_m/\mathcal{P},~& (nh)
		H_mK_x,~& (nh)
		\end{cases}
	    <\frac{H_mN}{|U|}<	
		\begin{cases}
		1,~& (h),\\
%		\sqrt{1+(2\pi H_m/\mathcal{P})^2},~& (nh).
		\sqrt{1+H_m^2K_x^2},~& (nh)
		\end{cases}\quad\text{at }Z=0.
	\end{align}
	Here, the lower bound originates from \eqref{eq:statio_kz}.
	It provides a real-valued and negative vertical wavenumber.
	The upper bound is due to \eqref{eq:stat_stab} 
	and guarantees static stability.
	\item modulationally stable if
	\begin{align}
		\frac{H_mN}{|U|}\neq\sqrt{2}\quad\text{at }Z=0.
	\end{align}
	This non-resonance criterion stems from the investigation of the essential spectrum in section \ref{sec:es}.
\end{itemize}
An illustration combining these criteria for the wave-Froude number 
computed by inversion of \eqref{eq:growavefr} is plotted in figure \ref{fig:wave_froude}.
It turns out that the resonance condition for the wave-Froude number is the same as the threshold 
that guarantees real-valued wave action in \eqref{eq:bca}.
Combing this result with the upper bound for real-valued wave action \eqref{eq:wavefroude}, we obtain a strict inequality,
\begin{align}
	\label{eq:wavefroude_bd}
	\mathit{Fr}_\mathrm{wave}<\frac{1}{\sqrt{2}},
\end{align}
for the existence and modulational stability of the wave.
That the critical values are the same does not come as a surprise because 
$A^+$ and $A^-$ coincide at this value
which is linked to the singularity of the Jacobian of the boundary condition by the
inverse function theorem.

Also, beyond $H_mK_x=\sqrt{2}$ waves cease to exist.
Only solutions where $H_mK_x>1$ may become modulationally unstable since waves 
in the region $H_mK_x<1$ overturn before they ever reach the resonance condition.

\section{Discussion}
\label{sec:conclusion}

The main result of this paper is that the stability of stationary strongly nonlinear gravity waves
with respect to Grimshaw's dissipative modulation equations depends on three characteristic parameters:
The wave-Reynolds number, the wave-Froude number and the mountain's height-period ratio.
Considering horizontally periodic waves from the surface to the deep atmosphere,
we find that the stability is completely determined by the boundary and far-field conditions.
These results are valid for fairly general wave solutions that posses only minor restrictions on the background fields:
the background is hydrostatic and exhibits a physical far-field behavior. 
Other than this, the background temperature and horizontal wind are unconditioned.

We want to give some useful remarks on the characteristic parameters.
When reformulated in dimensional variables, the wave-Reynolds number reads
\begin{align}
	\mathit{Re}_\mathrm{wave}=\frac{C_{gz}^\ast D^\ast}{\Lambda^\ast}
\end{align}
where $C_{gz}^\ast$ represents the vertical group velocity 
and $D^\ast=H_p^{\ast -1}|\pmb{K}^\ast|^{-2}$ defines a length scale 
with $H_p^\ast$ the local pressure scale height.

The net wave-Froude number as defined in this work is readily redimensionalized, so
\begin{align}
	\mathit{Fr}_\mathrm{wave}=\frac{H_m^\ast N^\ast}{|\bar{U}^\ast|}.
\end{align}
It has to be distinguished from the gross wave-Froude number
as the latter contains the induced mean-flow and is therefore not independent of the wave itself.
In the literature, only the gross wave-Froude number is considered which is reasonable since from observations one gets the mean-flow horizontal wind.
Generally, it is not feasible to ask for the background flow of a mountain wave
which really is the flow without the mountain.
However, knowing the wave parameters of the excited wave, it it possible to compute the background wind
by the total momentum equation \eqref{eq:degeneration}.
Also, in weakly nonlinear theory they are indeed the same
as the induced mean flow is a higher-order correction. 
Still, the strongly nonlinear description in this paper reduces the mean-flow wind and therefore
gains wave energy from the mean flow when excited. 
Hence, from a theoretical point of view the net wave-Froude number should not depend \textit{a priori} on the wave 
that is excited by the background flow over the mountain.

	\section*{Acknowledgments}
	This research was supported by the German Research Foundation (DFG) through Grants KL 611/25-2
of the Research Unit FOR1898 and Research Fellowship SCHL 2195/1-1.	
	
	\appendix

	%	\section{Insert appendix name here}
%	\label{app:weelpos}

% 	\clearpage

%	\printbibliography[heading=bibintoc]
	\bibliographystyle{abbrvnat}
	\bibliography{library.bib}

\end{document}